\documentstyle[epsfig,11pt]{article}

\def\pr#1#2#3{ Phys. Rev. ${\bf{#1}}$ (#2) #3}

\def\pl#1#2#3{ Phys. Lett. ${\bf{#1}}$ (#2) #3}

\def\np#1#2#3{ Nucl. Phys. ${\bf{#1}}$ (#2) #3}
\def\zp#1#2#3{ Z. f. Phys. ${\bf{#1}}$ (#2) #3}

\newcommand{\fig}[1]{Fig.\ref{#1}}
\newcommand{\tab}[1]{Table \ref{#1}}

\newcommand{\eqn}[1]{Eq.(\ref{#1})}
\newcommand{\eqns}[2]{Eqs.(\ref{#1},\ref{#2})}
\newcommand{\gev}{\mbox{ GeV}}

\newcommand{\ben}{\begin{enumerate}}
\newcommand{\een}{\end{enumerate}}
\newcommand{\bit}{\begin{itemize}}
\newcommand{\eit}{\end{itemize}}
\newcommand{\bc}{\begin{center}}
\newcommand{\ec}{\end{center}}
\newcommand{\bb}{\begin{bf}}
\newcommand{\eb}{\end{bf}}
\newcommand{\bsm}{\begin{small}}
\newcommand{\esm}{\end{small}}
\newcommand{\bns}{\begin{normalsize}}
\newcommand{\ens}{\end{normalsize}}
\newcommand{\bq}{\begin{equation}}
\newcommand{\eq}{\end{equation}}
\newcommand{\bqa}{\begin{eqnarray}}
\newcommand{\eqa}{\end{eqnarray}}

\newcommand{\vb}{\vspace*{2cm}}

\newcommand{\nn}{\nonumber}

\newcommand{\bft}{\begin{footnotesize}}
\newcommand{\eft}{\end{footnotesize}}

\def\awf{\alpha_{W\Phi}}
\def\abf{\alpha_{B\Phi}}
\def\aw{\alpha_{W}}

\def\dbw{\tilde{\alpha}_{BW}}
\def\dw{\tilde{\alpha}_{W}}
\def\tk{\tilde{\kappa}}
\def\tl{\tilde{\lambda}}

\def\b{{\cal B}}

\def\ctg{\mbox{ctg~}}

\def\c2{\chi^2}
\def\SM{Standard Model\ }

\def\L{{\cal L}}
\def\O{{\cal O}}
\def\OO{\tilde{{\cal O}}}

\def\vtau{\mbox{\boldmath $\tau$}}
\def\vW{\mbox{\boldmath $W$}}
\def\vvW{\tilde{\mbox{\boldmath $W$}}}
\def\kg{\kappa_\gamma}\def\kZ{\kappa_Z}
\def\xg{x_\gamma}\def\xZ{x_Z}
\def\dZ{\delta_Z}\def\lg{\lambda_\gamma}
\def\lZ{\lambda_Z}
\def\bqp{\bar{q}'}
\def\unp{${\cal P}_{LR}+{\cal P}_{RL}$}

\def\pup{${\cal P}_{RL}$}

\def\kirki{`$\kappa\acute{\iota}\varrho\kappa\eta$'\ }
\def\demo{$\Delta\eta\mu \acute{o} \kappa \varrho \iota \tau o \varsigma$}
\newcommand{\trsub}[1]{{_{_{{#1}}}}}

\def\fone{ \begin{figure}[!ht]
\begin{center}
\mbox{\epsfig{file=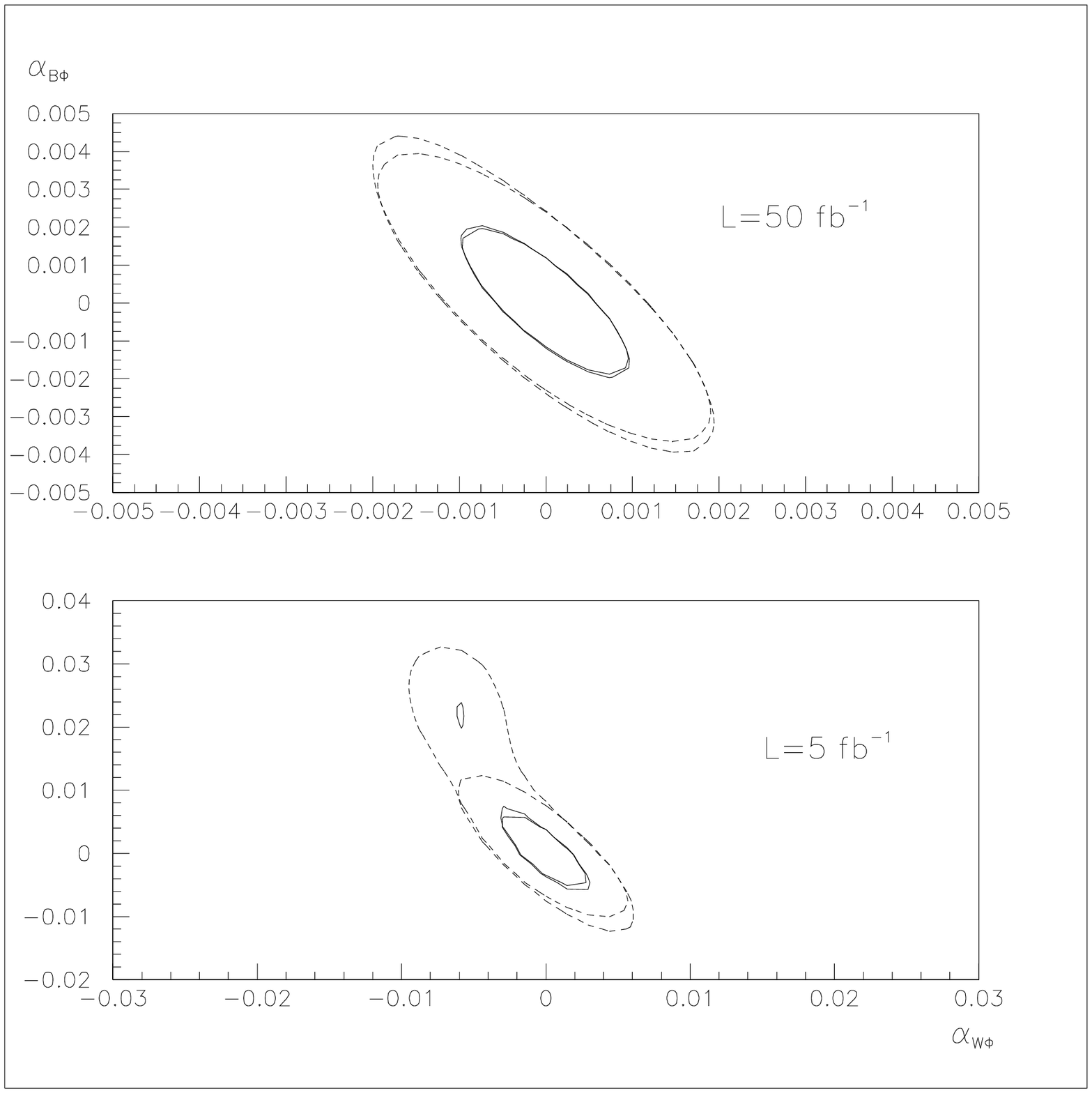,bbllx=1.5cm,bblly=0cm,bburx=18cm,bbury=27cm,height=13cm,width=9cm}}
\vspace*{-2cm}
\caption[.]{The one (solid) and two (dashed) standard 
deviations limits on $\abf$-$\awf$, using the optimal
observables and the EML methods for $\sqrt{s}=800$ GeV 
and unpolarized beams. In the upper part optimal observables and EML 
are hardly distinguishable, whereas in the lower part EML exhibits 
a secondary minimum.}
\label{comparison}
\end{center}
\end{figure} }
\def\ftwo{\begin{figure}[!ht]
\begin{center}
\mbox{\epsfig{file=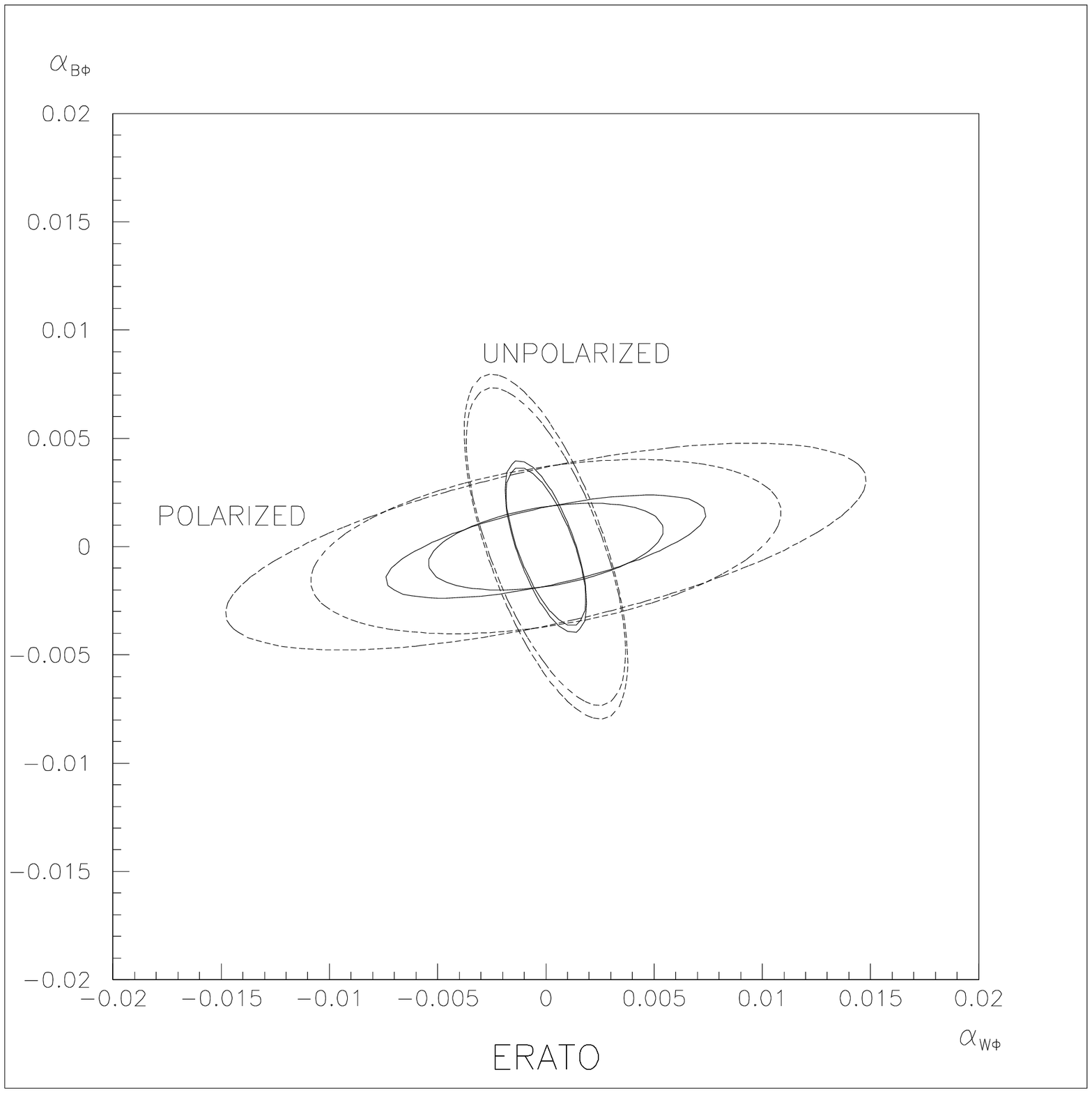,bbllx=1.5cm,bblly=0cm,bburx=18cm,bbury=27cm,height=13.5cm,width=9cm}}
\vspace*{-2cm}
\caption[.]{The one (solid) and two (dashed) standard deviations limits 
on $\abf$-$\awf$ for unpolarized and polarized $(e^-_Re^+_L)$ beams. 
The inner lines are from electron channel 
whereas the outer ones
are from muon channel.}
\label{polarization}
\end{center}
\end{figure}}

\begin{document}
\input feynman
 
\pagestyle{empty}
 
\begin{flushright}
DEMO-HEP-96/04 \\
THES-TP 96/11\\
{\tt hep-ph/9612378} \\
\end{flushright}
 
\vspace*{2cm}
\bc
{\LARGE\bf
Studying Trilinear Gauge Couplings \\[12pt]
at Linear Collider Energies 
}\\ 
\vspace*{2cm}

{\Large
G.J.~Gounaris$^a$$^1$ and C.~G.~Papadopoulos$^b$$^2$
} \\[12pt]
$^a$Department of Theoretical Physics, 
University of Thessaloniki, 54006 Thessaloniki, Greece
\\[0.5cm]
$^b$Institute of Nuclear Physics, NCSR \demo, 15310 Athens, Greece
\\
\vb
{\bf ABSTRACT}\\[12pt]  
\begin{quote}

We investigate the sensitivity of the semileptonic processes 
$e^+e^-\to\ell^-\bar{\nu}_\ell\; q\; \bqp$, $\ell=e\;\mbox{or}\;\mu$,
on the non-standard trilinear gauge couplings, using
the optimal observables method at Linear Collider energies.   
Our study is based on the four-fermion generator \verb+ERATO+.
Taking into account all possible 
correlations between the
different trilinear gauge coupling parameters, we show that 
they can be measured
with an accuracy of $10^{-3}$ to $10^{-4}$ for typical Linear Collider
energies and luminosities.

\end{quote}
\ec
\vspace*{\fill}
\noindent\rule[0.in]{4.5in}{.01in} \\      
\vspace{.3cm}  
E-mail: 
\parbox[t]{12.6cm}{ {$^1$gounaris@physics.auth.gr},
{$^2$Costas.Papadopoulos@cern.ch}.} 
\newpage
\pagestyle{plain}
\setcounter{page}{1}

\par
The future $e^+e^-$ linear colliders (LC), with energies ranging from
a few hundreds of GeV up to a couple of TeV, 
provide particle physics with an enormous potential for studying Nature
to the deepest level ever achieved~\cite{lc}. 
Although at these energies direct searches for new particles
and their interactions will eventually attract most of the 
physics interest, 
LC offer also the unique possibility to study to an extremely
high accuracy, the properties
of the existing particles, like those of the massive electroweak
gauge bosons $W$ and $Z$. 
Therefore, an important project at the LC energies
will be the determination of the trilinear gauge
couplings (TGC)~\cite{goun:0,hagi:0}, which are 
a characteristic manifestation of the underlying
non-Abelian symmetry of elementary particle 
interactions~\cite{sta-mod} and at the same time an interesting
probe of New Physics (NP).

 
\par
In order to study the TGC 
we need a parameterization of the vector gauge boson
interactions that goes beyond the \SM. The most 
general such parameterization is given by~\cite{lep2tgc}:
\bqa
\L_{TGC}&=&\sum_{V=\gamma,Z} -ie\: g_{\scriptscriptstyle VWW } 
\left( g_1^V ( V_\mu W^{-\mu\nu}W^{+}_{\nu}
-V_\mu W^{+\mu\nu}W^{-}_{\nu} )
+\kappa_V V_{\mu\nu}W^{+\mu}W^{-\nu}\right)\nn\\
&-&ie\;g_{\scriptscriptstyle VWW }\:{\lambda_V\over m_W^2}V_{\mu\rho}
W^{+\rho\nu}W^{-\mu}_\nu
\nn\\
&+&e\,g_{\scriptscriptstyle VWW} g_5^V \varepsilon_{\mu\nu\rho\sigma}
\left(
 (\partial^\rho W^{-\mu})W^{+\nu}
-(\partial^\rho W^{+\nu})W^{-\mu}
\right) V^\sigma  \nn\\
&+&e\,g_{\scriptscriptstyle VWW} \Bigg [ 
g_4^V W^+_\nu W^-_\mu \left(\partial^\mu V^{\nu}+\partial^\nu V^{\mu}\right)
+i \tilde{\kappa}_V W^+_\nu W^-_\mu {\cal V}^{\mu\nu}
+i {\tilde{\lambda}_V\over m_W^2} 
{W^{+\mu}}_\nu W^-_{\rho\mu}{\cal V}^{\nu\rho} \Bigg ]
\label{lagrangian}
\eqa
where
\[ V_{\mu\nu}=\partial_\mu V_{\nu}-\partial_\nu V_{\mu},\;\;
W^\pm_{\mu\nu}=\partial_\mu W^\pm_{\nu}-\partial_\nu W^\pm_{\mu},\;\;
\]
and 
\[ {\cal V}^{\mu\nu}=
\frac{1}{2}\varepsilon^{\mu\nu\rho\sigma} V_{\rho\sigma} .\;\;
\]
In \eqn{lagrangian} $W^{\pm}$ is the $W$-boson field, and
the usual definitions $g_{\scriptscriptstyle\gamma WW}=1$, 
$g_{\scriptscriptstyle ZWW}=\ctg\theta_w$
are used. In the \SM 
we have  $g_1^\gamma =g_1^Z=1$, $\kg=\kZ=1$, while  
all the other parameters are vanishing at tree level.
In searching for possible TGC, it is more convenient to express 
them in terms of their deviations from the \SM values. For this we
define the deviation parameters~\cite{lep2tgc,bile:0}:
\bq
\dZ=(g_1^Z-1) \ctg\theta_w\;\;\;\;
\xg=\kg-1\;\;\;\;
\xZ=(\kZ-1)\ctg\theta_w - \dZ \; \ \ ,
\label{deviation}
\eq
while we throughout assume $g_1^\gamma =1$, disregarding 
the possibility of an anomalous contribution to the
electromagnetic form factor of $W^\pm$.
We note that the NP contribution described by the
interaction Lagrangian in \eqn{lagrangian}, becomes linear when
expressed in terms of the above deviation parameters and $\lg$, $\lZ$, 
as well as the C- and P-violating couplings.
\par

During the last years, considerable progress has been achieved
concerning the understanding of the physics underlying
the non-standard boson self-couplings.
As showed in reference~\cite{goun:gauge},
the deviations from the \SM TGC couplings can be parameterized in a
manifestly gauge-invariant way by using the effective
Lagrangian approach and considering gauge-invariant operators 
involving higher-dimensional
interactions among the gauge bosons and the Higgs field. These operators
are scaled by an unknown parameter $\Lambda_{NP}$ describing 
the characteristic scale of some high energy New Physics,
generating at low energies the effective interaction 
$\L_{TGC}$  as a residual effect.
In order to generate all kinds of TGC appearing in \eqn{lagrangian},
we need operators with dimension up to twelve. On the other
hand, restricting ourselves to
$SU(2)_L\times U(1)_Y$-invariant operators with dimension six,
which are the lowest order ones in a $1/\Lambda_{NP}$ expansion,
we end up with the following list of operators capable to induce
TGC NP 
couplings~\cite{Buchmuller,hagi:1,deRujula1,hagi:2,Tsirigoti,Papadamou}:
\bqa
\O_{B\Phi}&=& i B^{\mu\nu}(D_\mu\Phi)^\dagger(D_\nu\Phi)\nn\\
\O_{W\Phi}&=&i (D_\mu\Phi)^\dagger\;\vtau\cdot\vW^{\mu\nu}
(D_\nu\Phi)\nn\\
\O_W&=&{1\over 3!}(\vW^{\mu}_{\;\rho}\times\vW^{\rho}_{\;\nu})
\cdot\vW^{\nu}_{\;\mu}
\label{operator}
\eqa
and\footnote{The most complete list of CP violating \verb+dim=6+ 
purely bosonic operators is given in~\cite{Tsirigoti}.
Concerning them we note that the TGC 
couplings generated by  $\OO_{BW}$, are 
identical to those induced by the operators $2 \OO_{B\Phi}/g$ or 
$2 \OO_{W\Phi}/g^\prime$, defined as the CP violating
analogs of  $ \O_{B\Phi}$ and $\O_{W\Phi}$ respectively.}
\bqa
\OO_{BW}&=&  \Phi^\dagger\;\frac{\vtau}{2} \cdot \vvW^{\mu\nu}
\Phi B_{\mu \nu}\nn\\
\OO_W&=&{1\over 3!}(\vW^{\mu}_{\;\rho}\times\vW^{\rho}_{\;\nu})
\cdot\vvW^{\nu}_{\;\mu} \ ,
\label{operator-cp}
\eqa
where
\bq 
\tilde{B}^{\mu\nu}=\frac{1}{2}\varepsilon^{\mu\nu\rho\sigma}
B_{\rho\sigma}\;\;
\ , \ \;\;\vvW^{\mu\nu}=\frac{1}{2}\varepsilon^{\mu\nu\rho\sigma} 
\vW_{\rho\sigma}\;.
\eq
In \eqn{operator} and \eqn{operator-cp}, 
$\tau_i$ describe the Pauli matrices,
\bqa
B_{\mu\nu}=\partial_\mu B_{\nu}-\partial_\nu B_{\mu}
\nn\eqa
is the $U(1)_Y$ gauge field strength,
\bqa
\vW_{\mu\nu}=\partial_\mu \vW_{\nu}-\partial_\nu \vW_{\mu}
-g\vW_{\mu}\times\vW_{\nu} \ \  \nn\eqa
is the field strength for  the $SU(2)_L$ gauge field $\vW_\mu$, and the 
Higgs doublet is written as 
\bqa
\Phi=\left( \begin{array}{c} \phi^+\\
{1\over \sqrt{2}}(v+H+i\phi^0)\end{array}\right) \ \ ,
\nn\eqa
while $D_\mu$ is the covariant derivative 
\bqa
D_\mu=\partial_\mu+i\; g\frac{\vtau}{2}
\cdot\vW_\mu+i\; g^\prime Y B_\mu \ \ ,
\nn\eqa
and  $Y$ is the hypercharge of the field on which $D_\mu$ acts.
Finally  $e=g\sin\theta_w=g^\prime\cos\theta_w$. \par

In the list of \eqns{operator}{operator-cp}, 
we have included all  \verb+dim=6+ purely bosonic 
operators 
contributing to the trilinear gauge interactions,
except those 
which give also a tree level contribition to LEP1 observables,
(and of course those which give no TGC at all).
This constitutes part of a consistent general strategy for searching
for any purely bosonic NP interaction. According to this
strategy, the measurement of TGC 
provides the most efficient way to
study the operators appearing in \eqns{operator}{operator-cp},
while the rest of \verb+dim=6+ operators can be most efficiently 
disentagled and
constrained either by high precision measurements (LEP1), or by
studying other production processes 
at LC~\cite{hagi:1,Verzegnassi,Tsirigoti}
and high-energy hadronic colliders\footnote{This is particularly needed for 
the gluon involving operators; (see~\cite{Papadamou}).}.

\par
The New Physics contribution from the above operators is
described by the effective Lagrangian 
\bqa
\L_{TGC}&=&
 g^\prime {\abf\over m_W^2}\O_{B\Phi}
+ g{\awf\over m_W^2}\O_{W\Phi}
+g{\aw\over m_W^2}\O_W \nn\\
&+& \frac{g g^\prime}{2} {\dbw\over m_W^2}\OO_{BW} 
+ g{\dw\over m_W^2}\OO_W 
\label{gi-lagrangian}
\eqa
where the relations between $\awf$, $\abf$, $\aw$, $\dbw$, $\dw$,  
and the deviation parameters of \eqn{deviation} are given by
\bqa
&\dZ=\awf/\left( \sin\theta_w \cos\theta_w\right)\;\;\;\;
\xg = \abf +\awf \;\;\;\;
\lg =  \aw &           \nn\\                        
&\xZ = -\tan\theta_w \xg   \;\;\;\;
\lZ = \lg& \nn\\           
&\tk_\gamma = \dbw \;\;\;\;
\tl_\gamma =  \dw &           
\nn\\  &\tk_Z = -\tan^2\theta_w \tk_\gamma   \;\;\;\;
\tl_Z = \tl_\gamma& \ \ .
\label{relation}
\eqa
As it is seen from \eqn{relation}, the restriction to New Physics
generated by \verb+dim=6+ gauge
invariant operators, implies that there are only five independent 
non-standard triple gauge couplings, three of which are 
CP-conserving~\cite{schild} and two CP-violating.



In order to study the effect of TGC, one usually considers the
reaction $e^+ e^-\to W^+ W^-$, taking into account the subsequent
decay of the two $W$'s to a four-fermion final state~\cite{bile:0}. 
Such final states can be classified in 
three categories, namely
the `leptonic' 
$\ell_1^- \bar{\nu}_{\ell_1} \ell_2^+ \nu_{\ell_2}$,
the `semileptonic' $\ell^-\bar{\nu}_\ell\; q \bqp$ and the
'hadronic' channel $ q_1 \bqp_1 \bar{q}_2 q^\prime_2$, 
(where $q$ and $q^\prime$ refer to up- and down-type quarks respectively). 
Semileptonic seems to be the most favoured 
channel~\cite{lep2tgc} for 
studying TGC, since
it contains the maximum kinematical information; taking into account that
charge-flavour identification in a four jet channel is rather
inefficient and that
the cross section for the leptonic mode is suppressed.
Thus, in the present paper we 
study at LC energies, the TGC effect induced by the interaction 
\eqn{gi-lagrangian} in the processes 
\bq 
e^+e^-\to\ell^-\bar{\nu}_\ell\; q \bqp 
\label{processes}
\eq
where $\ell$ is an electron or a muon. 
The final state 
$\tau \bar{\nu}_\tau\; q \bqp$ is
not  considered here, due to the special difficulties to identify
$\tau$'s in this environment.

\begin{figure}[ht]
\unitlength=0.01pt
\bc
\begin{picture}(10000,12000)
\global\Xone=5000
\global\Xtwo=2500
\global\Yone=3
\global\Ytwo=4

\pfrontx = 0
\pfronty = 2000

\drawline\fermion[\E\REG](\pfrontx,\pfronty)[10000]
\global\advance \pfrontx by -1000 
\put(\pfrontx,\pfronty){$e^+$}
\global\advance \pbackx by 1000 
\put(\pbackx,\pbacky){$\bar{\nu}_e$}
\global\gaplength=250
\drawline\scalar[\N\REG](\pmidx,\pmidy)[3]
\global\advance \pmidx by 500 
\put(\pmidx,\pmidy){$W$}

\drawline\photon[\N\REG](\pbackx,\pbacky)[\Ytwo]
\global\advance \pmidx by 500 
\put(\pmidx,\pmidy){{$\gamma,Z$}}
\global\gaplength=250
\drawline\scalar[\E\REG](\pfrontx,\pfronty)[3]
\drawline\fermion[\NE\REG](\pbackx,\pbacky)[\Xtwo]
\global\advance \pbackx by 500 
\put(\pbackx,\pbacky){$q$}
\drawline\fermion[\SE\REG](\pfrontx,\pfronty)[\Xtwo]
\global\advance \pbackx by 500 
\put(\pbackx,\pbacky){$\bar{q}'$}

\drawline\fermion[\E\REG](\photonbackx,\photonbacky)[5000]
\global\advance \pbackx by 500 
\put(\pbackx,\pbacky){$e^-$}
\drawline\fermion[\W\REG](\pfrontx,\pfronty)[5000]
\global\advance \pbackx by -1000 
\put(\pbackx,\pbacky){$e^-$}

\end{picture}
\ec
\caption[.]{Single-resonant graph where TGC are contributing.}
\label{fig0}
\end{figure}
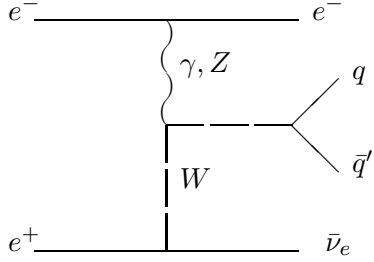

\par Quite often, the four-fermion final state processes, 
\eqn{processes}, are calculated by just taking into account
contributions from the
$e^+ e^-\to W^+ W^-$ subprocess, which is equivalent to a narrow
width approximation $\Gamma_W\to 0$. In the classification
of the four-fermion production diagrams of 
reference~\cite{lep2gen}, these graphs are termed as 
the double-resonant graphs \verb+CC03+. 
Such a narrow width approximation neglects contributions from 
single-resonant graphs, which become increasingly important
at higher energies, (at least in some parts of the phase 
space)~\cite{lep2gen,pap:4f,berends-an,gintner}.
The situation is particularly severe for final states involving
$e^\pm$, where graphs like the one presented in
\fig{fig0}, which involves a $t$-channel
photon exchange, dominate in certain parts of the phase-space
at higher energies. Moreover, the graph of \fig{fig0} receives contributions
from the trilinear gauge boson interactions which are not included
in the $e^+ e^-\to W^+ W^-$ calculation. In order to perform an analysis,
as complete as possible, it is therefore mandatory to include in the
calculation of the processes \eqn{processes} all tree-order diagrams,
resonant as well as non-resonant ones.\par

Nowadays this is possible, since there exist widely available  
four-fermion codes, where the TGC effects are included beyond the 
narrow width approximation~\cite{lep2tgc,pap:4f,berends-an}. 
In the calculations presented in this paper we have used 
for this purpose, the \verb+ERATO+ Monte-Carlo event generator 
described in~\cite{lep2gen,pap:4f,erato}. The basic ingredients of this 
generator are the following:
\ben
\item Exact tree-order helicity amplitudes for the processes 
$ e^+e^-\to\ell^-\bar{\nu}_\ell\; q \bqp$,  
including all trilinear gauge interactions described by
\eqn{lagrangian}~\cite{pap:4f,pap:old}.
\item Phase-space generation algorithm based on a multi-channel 
Monte Carlo approach, including 
weight optimization~\cite{weightopt}.
\item Treatment of the unstable particles contribution
consistent with gauge-invariance and high-energy 
unitarity~\cite{pap:4f,bhf,bhf2}.
\item Initial state radiation (ISR), based on the structure function
approach~\cite{lep2ww}, including soft-photon exponentiation 
as well as hard collinear photon emission in the leading logarithmic 
(LL) approximation up to order $O(\alpha^2)$.
\item Coulomb correction\footnote{For a detailed description see 
reference~\cite{lep2ww}.} 
to the double resonant (\verb+CC03+) graphs.
\item Beamstrahlung effects have also been included via the \kirki 
algorithm~\cite{circe}.
\een

Apart from the beamstrahlung effects just mentioned,
the treatment of the higher order corrections
in the present study is the same as in the LEP2 case.

\par 
In order to avoid matrix element singularities and to 
be as close as possible to the experimental situation, 
we have applied the cuts
\bq
175^o  \geq ( \theta_\ell\;,\;\theta_{jet} ) \geq 5^o
,\;\;\;
E_\ell \geq  10 \gev\;,\;\; E_{jet} \geq 10 \gev\;\;\mbox{and}\;\;
m_{q,\bqp}\geq 15 \gev \ .
\label{cuts}
\eq 
Finally, we use the Standard Model input parameters 
\bqa 
&M_W = 80.23\gev,\;\;\Gamma_W=2.033\gev,\;\;
M_Z=91.188\gev,\;\;\Gamma_Z=2.4974\gev,&\nn \\
&\sin^2\theta_w = 0.23103\;\;\mbox{and}\;\; 
\alpha(M_Z)=1/128.07 \ , &
\label{inputs}
\eqa
while in the ISR structure function $\alpha=1/137.036$
is of course used. For the analysis of the
beamstrahlung effects we have used the \verb+TESLA+ design. 


\par 
In order to determine the sensitivity of
a given reaction on the TGC parameters one has to 
maximize the likelihood function~\cite{sekulin}, whose logarithm is 
given by
\bq \ln \L_{ML}=\sum_{i}^{N} \ln p(\Omega_i,\vec{a})\;\;,
\label{likelihood}
\eq 
where the sum is running over the event sample under investigation.
$\Omega_i$ represents the collection of the independent 
kinematical variables describing the i-th event, 
the vector $\vec{a}$  is defined in the coupling space as 
$\vec{a}=(\awf, \abf, \aw, \dbw, \dw)$, and
\bq 
\label{probability}
p(\Omega_i,\vec{a})={1\over \sigma}{d\sigma\over d\Omega}
\Big|_{|\Omega=\Omega_i}
 \ \ \ , 
\eq
\bq 
\sigma=\int_{V} {d\sigma\over d\Omega} d\Omega\;\; ,
\eq
is the probability to find an event at the phase-space point
$\Omega_i$.
Since 
the interaction Lagrangian is linear with respect to the
TGC parameters, one can  write the differential cross section 
in the form
\bq 
{d\sigma \over d\Omega}=c_0(\Omega)+\sum_{i}a_i c_{1,i}(\Omega)
+ \sum_{i,j}a_i a_j c_{2,ij}(\Omega)
\label{differentialxs}
\eq
and similarly the total cross section as 
\bq 
\sigma=\hat{c}_0+\sum_{i}a_i \hat{c}_{1,i}
+ \sum_{i,j} a_i  a_j \hat{c}_{2,ij} \ ,
\eq
where hatted $c$'s are integrals of unhatted ones over the phase space.


\par 
The sensitivity on the TGC parameters is determined by the so-called
information matrix~\cite{eadi} given by the second derivative of the
likelihood function,
\bqa 
I_{ij}& \equiv &E\bigg[ \left({\partial\over \partial  a_i} \ln\L_{ML}\right)
\left({\partial\over \partial  a_j} \ln\L_{ML}\right) \bigg] \\
&=& - E\bigg[ {\partial\over \partial  a_i}{\partial\over \partial  a_j}
\ln\L_{ML}\bigg]\nn
\label{infomatrix}
\eqa
where
\bq 
E[ A] = \int \prod_{i=1}^{N} \{ d\Omega_i\;p_0(\Omega_i)\} 
A(\Omega_1,...,\Omega_N)
\eq
represents the mean value of a function $A(\Omega_1,...,\Omega_N)$.
If we assume that the maximum of the likelihood function 
is located at $\vec{a}=0$,
which reflects the physical expectation that the `data' will be 
consistent with  the \SM, 
the information matrix is given to the lowest order, 
by $I_{ij}=N \b_{ij}$, with
\bq 
\b_{ij}\equiv  \left\langle { c_{1,i}\over c_0} { c_{1,j}\over c_0} 
\right\rangle_0
- \left\langle { c_{1,i}\over c_0}\right\rangle_0 
  \left\langle { c_{1,j}\over c_0}\right\rangle_0
\eq
and 
\bq
\label{A0average}
\langle A \rangle\trsub{0} = \int d{\Omega}\; p_0(\Omega)\; 
A(\Omega)\;\; ,
\eq
\bq 
\label{probability0}
p_0(\Omega)={1\over \sigma}{d\sigma\over 
d\Omega}\bigg|_{\vec{a}=0}\;\; .
\eq

In the optimal observables approach~\cite{optobs},
equivalent results are obtained by defining the  phase-space functions
\bq
\O_i= { c_{1,i}(\Omega)\over c_0(\Omega)} \;\;
\label{optimal}
\eq
called optimal observables, whose mean values and 
covariance matrix determine the sensitivity on the TGC 
parameters.
More specifically one writes
\[
\bar{ a}_i=\sum_j \b^{-1}_{ij} \left( \langle \O_j \rangle 
- \langle \O_j \rangle\trsub{0}
\right)
\]
as an unbiased estimator of the components of  $\vec{a}$, 
while the corresponding covariance matrix is given by
\[
V(\bar{ a} )= {1\over N}\;\; \b^{-1} \cdot V(\O ) \cdot \b^{-1}
\]
with $V(\O )$ defined by
\[
V(\O)_{ij} =   \langle \O_i \O_j \rangle -
 \langle \O_i\rangle \langle \O_j\rangle \; , 
\]
and
\[
\label{Aaverage}
\langle A \rangle = \int d{\Omega}\; p(\Omega)\; 
A(\Omega)\;\; .
\]
Under the assumption that the `data' are accounted for by 
the Standard Model,
we have that $V(\O)=\b$, which shows that to the lowest order,
the likelihood approach and the optimal observables are indeed
equivalent.

\par 
Up to now we have considered the so called Maximum Likelihood method.
One can improve the analysis by considering also the so called Extended
Maximum Likelihood (EML). In this case we take into account 
the variation of the total number of expected events as a function
of the unknown parameters and define 
\bq 
\L_{EML}=p_N \prod_{i}^{N}  p(\Omega_i,\vec{a})\;\;,
\eq 
where 
\[
p_N= \frac{ \langle N \rangle^N }{N!}
e^{-\langle N \rangle} \ 
\]
and $\langle N \rangle$ is the number of expected events.
The analysis proceeds as before
and  the correlation matrix is expressed as 
\bq
\b_{i,j}= \left\langle { c_{1,i}\over c_0} 
{ c_{1,j}\over c_0} 
\right\rangle_0
\eq
while the information matrix becomes $I_{ij}=\langle N  \rangle\b_{i,j}$.
The corresponding optimal observables in the EML approach are 
\bq
\O_i= N\;{ c_{1,i}(\Omega)\over c_0(\Omega)} \;\;.
\label{optimale}
\eq
 
\par 
In the sequel we perform one- as well as five-dimensional investigations.
One-dimensional investigations assume that all but one of the $a_i$'s
are non-vanishing, and lead to  parameter errors (1sd) given by
\bq 
\delta a_i={1\over\sqrt{N b_{ii} } } \;\;.
\eq
The multidimensional case where all five NP
couplings are considered simultaneously, is also treated. 
Diagonalizing then the symmetric correlation matrix $\b$, 
one first
determines its eigenvalues $\lambda_i$ and eigenvectors 
$\vec{e}_i$, normalized so that $\vec{e}_i\cdot\vec{e}_j=\delta_{ij}$.
For each $\vec{e}_i$, the parameter
\bq 
 a^D_i  \equiv \vec{e}_i\cdot\vec{a} \;\; 
\label{orthogonal} 
\eq 
is then defined, for which the (1sd) error is given by 
\bq 
\delta a^D_i={1\over\sqrt{ N \lambda_i}} \ . 
\label{orthogonalerrors} 
\eq

\par 
In all calculations presented in this paper, $N$ is taken to be 
the predicted \SM number of events defined by $N=4\;L\;\sigma_{SM}$,
where $\sigma_{SM}$ is the corresponding total cross section, 
$L$ is the integrated luminosity and the factor $4$ takes into account 
the four equivalent channels described by the same matrix elements; 
i.e. $ e^+e^-\to \ell^-\bar{\nu}_\ell\; u \bar{d}$,
$\ell^-\bar{\nu}_\ell\; c \bar{s}$, $\ell^+{\nu}_\ell\; d \bar{u}$
and $\ell^+{\nu}_\ell\; s \bar{c}$.

\fone

\par 
At this point we address the question, 
how accurately the optimal observables
approximation describes the ML (or EML) function. 
To this end we calculate the $\ln \L_{ML}$ 
by replacing the sum appearing in
\eqn{likelihood} by an
integral over the expected probability distribution which is assumed to be
the one predicted by the \SM
\bq  \ln \L_{ML}  =
N \int \prod_{i=1}^{N} \{ d\Omega_i\;p_0(\Omega_i)\} 
\ln p(\Omega_i,\vec{a})\;\;.
\eq 
It is  clear that optimal observables and ML methods become identical 
in the limit $N\to \infty$, since then only the
leading term in the expansion of the likelihood function
survives; and this
is exactly the term proportional to the information matrix.
On the other hand for relatively low statistics, the nonlinearity of the
likelihood function becomes important and the optimal
observables approximation breaks down.
These features are shown in \fig{comparison}, where the
one, and two standard deviation limits on $(\abf, ~\awf)$ 
are considered, for the muon channel and  
$\sqrt{s}=800$ GeV. In the upper part of the figure
the value of the integrated luminosity is taken to be 
$L=50\; \mbox{fb}^{-1}$, which is 
the expected nominal value, whereas in 
the lower part a much lower luminosity,  
$L=5\; \mbox{fb}^{-1}$, has been used.   

We have checked that for all nominal LC energies and luminosities,
the optimal observables
approximation gives identical results to those of the conventional
likelihood approach.
Furthermore the fact that the expected sensitivities
on the TGC parameters are predominantly determined by the
linear terms in the expansion of the differential cross section,
\eqn{differentialxs},
shows the self-consistency of our original
assumption that a parameterization of TGC in terms of
\verb+dim=6+ operators should be adequate.

On the other hand, from the point of view of a weighted Monte-Carlo
approach, which is frequently used in the phenomenological analyses,
the optimal observables method offers a very efficient 
 fast and economic way
to estimate not only the sensitivity of a given 
process on a single TGC parameter (or any kind
of `deviation' parameter), but also their full covariance matrix, which is
of great importance for multiparameter analyses.

Finally we would like to mention that 
the general
experimental problem of how to overcome the ISR as well as the detector
resolution induced
difficulties in the reconstruction of the event kinematics,  
is not addressed here.
We only note that this problem
exists also for the current LEP2 experiments
and that detailed experimental simulations at linear collider energies
can be found in the proceedings of the DESY-ECFA Workshop on
Linear Colliders~\cite{DESY-ECFA}.
Moreover our experience from LEP2 studies~\cite{lep2tgc}
shows that, despite the abovementioned problem, a very good estimate
of the sensitivity on the TGC can be obtained by an analysis
of the kind used in our present study.


\begin{table}[htb]
\bc
\begin{small}
\begin{tabular}{|c|rrrrr|rrrrr|}
\hline 
 & \multicolumn{5}{c|}{\unp} & \multicolumn{5}{c|}{\pup} \\ 
\hline
\hline & && && && && & \\ 
$\; e\;$ &
14.53  & 4.87  & 3.26 &  -0.0032 & -0.27 & 9.66 & -10.20&   0.33 &  0.59&  0.027\\
& 4.87  &  3.67 &  0.33&  0.016 & -0.072& -10.20 & 70.33 &-0.11 &-1.80 & -0.096\\
&3.26   &  0.33 & 20.75& -0.011 &  0.38 & 0.33 & -0.11 & 3.50 & 0.012 &-0.016\\
&-0.0032& 0.016 & -0.01& 0.28 &  -0.78&  0.59 & -1.80 & 0.012& 6.66 &   0.22\\
&-0.27  & -0.072&  0.38& -0.78 &  21.53 & 0.027& -0.096& -0.016& 0.22 &   3.46\\  
&& && && && &  &  \\  \cline{2-11} 
& \multicolumn{5}{c|}{456(5)} &  \multicolumn{5}{c|}{207(5)}  \\  
\hline \hline  & && && && && &   \\
$\;\mu\;$ 
& 
23.33  & 7.33  & 5.66 &  0.013 & -0.060 & 258.42 & -513.87 &  12.94 & -1.35 & -0.12\\
& 7.33  &  5.25 & 0.71&  0.033 & -0.054 &-513.87 & 2478.22 & -9.50 & 12.36 & 0.28\\
&5.66   &  0.71 & 33.00& -0.032 &  0.29 & 12.94 & -9.50 &2.71 & 0.27 &-0.0013\\
&0.013& 0.033 & -0.032& 0.28 &  -1.33   &  -1.35 &12.36 & 0.27& 178.16 &  12.11\\
&-0.060  & -0.054&  0.29& -1.33 &  33.85 & -0.12& 0.28 & -0.0013& 12.11&   2.67\\
&& && && && &  &  \\  \cline{2-11}
& \multicolumn{5}{c|}{267(2)} &  \multicolumn{5}{c|}{6.01(4)}  \\  
\hline 
\end{tabular}
\end{small}
\caption[.]{The correlation matrix for $e$ and $\mu$ channels at 500 GeV 
for the TGC parameters $\awf,\abf,\aw, \dbw, \dw $. Also shown
are the cross sections as well as their Monte Carlo errors in 
femptobarns. Here \pup (${\cal P}_{LR}$) corresponds to $e^-_Re^+_L$ 
($e^-_Le^+_R$) initial state polarization.} 
\label{tab-c}
\ec
\end{table}


In \tab{tab-c} we present the results for the correlation matrix 
$\b_{ij}$ involving all CP-conserving and CP-violating
couplings, at 500 GeV center of mass energy. 
The total cross sections are also presented.
As is evident from this table, the correlations between the different
$a_i$'s are not negligible in general, which suggests that 
 an analysis taking them into
account, is indispensable. 
\par 
An other very interesting result, is that electron and muon channels
exhibit a complementary behaviour:
electron channel gives the highest production rate, which means
a better statistics, whereas the muon-channel exhibits a higher
sensitivity on TGC.

\par
In \tab{tab-e} we show the eigenvalues of the correlation matrix, 
as well as the corresponding eigenvectors. These eigenvectors define 
directions in the five-parameter space, which are uncorrelated,
so that parameter errors can be safely extracted.
This table shows that for the unpolarized beams case, 
the dominant eigenvalues  correspond to directions 
in the five-parameter space related predominantly to $\aw$, $\awf$,
and $\dw$,
whereas the lowest ones are related to $\abf$ and $\dbw$.
The picture becomes almost opposite in the case that the
electron (positron) beams are right- (left)-polarized.

\begin{table}[htb]
\tabcolsep=3pt
\bc
\begin{small}
\begin{tabular}{|c|lrrrrr|lrrrrr|}
\hline 
 & \multicolumn{6}{c|}{$e$} & \multicolumn{6}{c|}{$\mu$} \\ 
\hline
\hline &&& && && && && & \\ 
${\cal P}_{LR}$
&
22.49 & -0.425 & -0.124 & -0.868 & 0.008 & -0.220 & 
36.10 & -0.460 & -0.129 & -0.872 & 0.004 & -0.099 \\
$+$
&21.54 &0.149 & 0.0477 & 0.167 & 0.035 & -0.972 & 
33.90 & 0.003 & 0.002 & -0.007 &  0.039 & -0.999 \\
${\cal P}_{RL}$
&14.71 & 0.813 & 0.344 & -0.463 & -0.002 &  0.061 &
22.88 & 0.816 & 0.320 & -0.479 & 0.0005 & 0.018\\
&1.77 & -0.366 & 0.929 & 0.046 &  0.011 & -0.002 &
2.60 &  -0.343 & 0.938 & 0.041 &  0.010 & 0.001 \\
&0.25 & 0.004 &  -0.010 &  -0.0006 & 0.999 & 0.036 &
0.22 & 0.002 &  -0.010 &  0.0007 &  0.999 &  0.039\\  
&& && && && &  &&  &  \\ 
\hline \hline  & &&&& && && && &   \\
\pup 
& 
72.06 & 0.161 & -0.986 & 0.002 &  0.028 & 0.001 & 
2591.53 & 0.215 &  -0.976 &  0.004 &  -0.005 &  -0.0001 \\
&8.07 & -0.963 & -0.163 & -0.067 &  -0.198 & -0.011 &
178.98 & -0.040 &  -0.003 &  -0.004 &  -0.996 &  -0.068\\
&6.56 & -0.199 & -0.0043 & -0.018 & 0.977 &  0.068 & 
145.96 & 0.973 &  0.214 &  0.073 &  -0.040 &  -0.003 \\
&3.48 & 0.062 &  0.007 &  -0.894 & -0.034 &  0.440 & 
1.87 & 0.068 &  0.010 & -0.935 & -0.022 &  0.346\\
&3.44 & -0.028 &  -0.003 &  0.440 &  -0.059 &  0.895 & 
1.83 & -0.024 &  -0.003 &  0.346 &  -0.064 &  0.935\\
&& && && && &&&  &   \\ 
\hline 
\end{tabular}
\end{small}
\caption[.]{The eigenvalues (2nd and 8th columns) 
and the corresponding eigenvectors of
the correlation matrices given in \tab{tab-c}.}
\label{tab-e}
\ec
\end{table}

\ftwo

As far as the polarization is concerned, we see that passing from
unpolarized to right-polarized
electron-beam, results to a much higher sensitivity for $\abf$
and $\dbw$ couplings.
These phenomena are much more pronounced for the muon-channel. 
This effect, which has been also 
observed in on shell studies of 
$e^-e^+ \to W^-W^+$~\cite{eeWWpol}, 
reflects the fact that different TGC parameters contribute to different 
helicity amplitudes, especially in the high-energy regime.
In \fig{polarization} we show how the information from both polarizations
can, in principle, be used to disentangle different TGC parameters, 
based on the fact that the corresponding covariance matrices are
very different. 

\par 
Finally in \tab{tab-bound}, one-standard-deviation errors 
are presented by combining electrons and muons as
\[
\b_{ij}=\b^{(e)}_{ij}{\sigma^{(e)}\over \sigma^{(e)}+\sigma^{(\mu)} }
+\b^{(\mu)}_{ij}{ \sigma^{(\mu)} \over \sigma^{(e)}+\sigma^{(\mu)} }\;\;,
\]
\[ N=4 L ( \sigma^{(e)}+\sigma^{(\mu)} )\;\;. \]
In our studies, $L$ is taken to be 20 fb$^{-1}$ at 500 GeV, 
10 fb$^{-1}$ at 360 GeV and
50 fb$^{-1}$ at 800 GeV. 
For the results concerning polarized beam scattering, we used  
\[
L_{\mbox{polarized}}=\frac{1}{4} \; L_{\mbox{unpolarized}}\;.
\]
In order to study the effect of the correlations among the TGC parameters
we distinguish two cases:
\begin{itemize}
\item The $1d$-case is based on the very strong and often  made
assumption that only one of the TGC parameters ($\awf$, $\abf$, $\aw$, 
$\dbw$, $\dw$) is non vanishing. A very contrived form of NP at
high energy scale is needed, in order to create such a situation where
only one of the operators appearing in \eqns{operator}{operator-cp}
is generated at low energies \cite{dyn}.  This case  
corresponds to the `one-dimensional log-likelihood fit'.
\item  
In the $5d$-case, on the contrary, the errors are calculated 
according to \eqn{orthogonalerrors}, where 
the full correlation matrix is taken into account and no {\it a priori} 
assumption on the size of the parameters has been made. 
Although in this case the presented errors correspond to directions
in the five-parameter space defined by~\eqn{orthogonal}, 
which are not generally identical to the ones defined
by the original parameters, we kept the same notation, since the
former are rather close to the latter: for instance $\awf^D$  
is mainly composed by $\awf$ and so on for the other 
TGC parameters~\cite{papa:optobs-lep2}. 
\end{itemize}
As it is seen from \tab{tab-bound}, moving from
the $1d$-case to the $5d$-case, 
the change on the one-standard-deviation errors 
reach the level of 40\%. Moreover the less sensitive the TGC parameter is,
the more the correlations affect its error. It should be
mentioned however that the correlations 
among the different TGC parameters do not dramatically change
the order-of-magnitude
estimate of their sensitivity based on single-parameter considerations.

Concerning the CP violating interactions, we should  note
that in reference ~\cite{deRujula}, bounds on the CP-violating
couplings $\tk_\gamma$ and $\tl_\gamma$ have been derived, on
the basis of 
their contribution to the electric dipole moment (EDM) of the meutron.
Besides the fact that these bounds depend on several details, 
the most they  imply is a
strong linear relation between $\tk_\gamma$ and $\tl_\gamma$. 
Direct measurements of these couplings, as well as of their
$Z$-boson counterparts,  will therefore be useful because they
will provide detail information on the whole 
CP-violating TGC parameter-space.

\begin{table}[htb]
\bc
\begin{tabular}{|c||c|cc|c|}
\hline 
$\sqrt{s}$ (GeV) & 360 & \multicolumn{2}{|c|}{500} & 800 \\
& 1d & 1d & 5d & 1d\\ \hline\hline 
$\awf$ & 0.0018 & 0.00098 & 0.00098 & 0.00042\\
       &         & \bft (0.0037)\eft  &  \bft(0.0045) \eft   & \\
\hline 
$\abf$ & 0.0039 & 0.0020  & 0.0028  &  0.00083\\
       &         & \bft (0.0013)\eft   &  \bft(0.0012)\eft  & \\
\hline 
$\aw $ & 0.0016  & 0.00082 & 0.00081 & 0.00031 \\
       &         & \bft (0.0082)\eft   &  \bft(0.0082) \eft  &   \\
\hline 
$\dbw$ & 0.011   & 0.0078  & 0.0084  & 0.0048 \\
       &         & \bft (0.0045)\eft   &  \bft(0.0044)\eft   & \\
\hline 
$\dw$  & 0.0016  & 0.00081 & 0.00079 & 0.00031 \\
       &         & \bft (0.0082)\eft   &  \bft(0.0083) \eft  & 
       
\\  \hline 
\end{tabular}
\caption[.]{One standard deviation errors on TGC parameters.
At 500 GeV we show also the effect of the correlations between different 
five TGC parameters in the $5d$ case for unpolarized beams, 
whereas in parentheses we show the corresponding errors from 
a right-handed polarized (left-handed) electron (positron) beam,
as explained in the text.}
\label{tab-bound}
\ec
\end{table}
\par
In studying the sensitivity on the TGC one
usually neglects possible correlations with other electroweak parameters
like for instance the non-standard
contributions to $Vf\bar{f}$ vertices, where $V$ stands for $Z$ or $W$. 
This is rather well justified
because the latter are usually much more constrained  
than the former as it is indeed the case at LEP2, where
TGC determination is expected to reach the level of 0.01 to 0.1,
compared with the constraints on the $Vf\bar f$ couplings
coming mainly from LEP1 analysis, which are at the level of 
10$^{-3}$~\cite{gual:leplatest}.
On the other hand, as it is also suggested by our analysis,
at LC energies the TGC can be tested
to a precision much higher than that of LEP2, and therefore
it becomes interesting to study the correlations 
among the TGC and the other electroweak couplings~\cite{casalbuoni,future}. 

Finally we would like to mention that a detailed comparison
of our study with those presented in references~\cite{bile:0,eeWWpol}
based on the on-shell production $e^- e^+ \to W^-W^+$
is not possible, due mainly to the fact that we are using different
analysis methods. Nevertheless both approaches agree in the order-of-magnitude
estimate of the expected sensitivity on the TGC.



\vspace*{1cm}

We conclude by summarizing the results of our study:
\ben
\item We have presented a five-parameter description of the 
      non-standard trilinear 
      gauge couplings, which includes all gauge-invariant contributions
      to the lowest order. We then analysed their contribution to the
      semileptonic reactions $e^+e^-\to \ell^-\bar{\nu}_\ell\; q \bqp $ 
      for $\ell=e$ and $\ell=\mu$ at LC energies, and showed that 
      a measurement of all five parameters is possible, with a sensitivity 
      covering a rather wide range starting from
      $1.1\times 10^{-2}$ (1sd) at 360 GeV for $\dbw$, (worst case),
      and going down to $3\times 10^{-4}$ (1sd) 
      at 800 GeV for $\aw$ and $\dw$, (best case).
\item The electron channel, due to the onset of the 
      single-$W$ production mode,
      gives the dominant contribution to the total 
      cross section at LC energies,
      whereas the muon channel exhibits a higher sensitivity on the 
      TGC parameters.  Therefore their overall
      contribution to the error on the TGC determination become equally 
      important. 
\item Polarization effects are important in order to disentangle different
      TGC contributions, leading to a substantial improvement
      of the sensitivity on the TGC parameters, especially for 
      $\abf$ and $\dbw$.   

\een

\vspace*{1cm}
\noindent {\bf Acknowledgements} \\[12pt]
We would like to thank the organizers as well as all the participants
of the ECFA-DESY Workshop on Linear Collider (1996), for their helpful
suggestions.
C.G.P, would like to thank the Department of Physics of the University of 
Thessaloniki, where part of this work was done, for its kind hospitality.
This work was partially supported by the EU grant CHRX-CT93-0319 and by the 
General Secretariat for Research and Technology ($\Gamma\Gamma\mbox{ET}$)
grant $\Pi\mbox{ENE}\Delta$-1995-350.

\end{document}